\documentclass[conference]{IEEEtran}
\IEEEoverridecommandlockouts
\usepackage{cite}
\usepackage{amsmath,amssymb,amsfonts}
\usepackage{algorithmic}
\usepackage{graphicx}
\usepackage{subcaption}
\usepackage{textcomp}
\usepackage{xcolor}

\def\BibTeX{{\rm B\kern-.05em{\sc i\kern-.025em b}\kern-.08em
    T\kern-.1667em\lower.7ex\hbox{E}\kern-.125emX}}
\begin{document}

\title{GNN-GMVO: Graph Neural Networks for Optimizing Gross Merchandise Value in Similar Item Recommendation\\
}

\author{\IEEEauthorblockN{Anonymous}

\author{\IEEEauthorblockN{ Ramin Giahi, Reza Yousefi Maragheh, Nima Farrokhsiar, Jianpeng Xu,  \\
Jason Cho, Evren Korpeoglu, Sushant Kumar, Kannan Achan}
\IEEEauthorblockA{\textit{Personalization Team} \\
\textit{Walmart Global Tech}\\
Sunnyvale, CA, USA \\
\{ramin.giahi, reza.yousefimaragheh, nima.farrokhsiar, jianpeng.xu, \\ jason.cho, ekorpeoglu, sushant.kumar, kannan.achan\}@walmart.com}
}

}

\maketitle

\begin{abstract}

Similar item recommendation is a critical task in the e-Commerce industry, which helps customers explore similar and relevant alternatives based on their interested products. Despite the traditional machine learning models, Graph Neural Networks (GNNs), by design, can understand complex relations like similarity between products. However, in contrast to their wide usage in retrieval tasks and their focus on optimizing the relevance, the current GNN architectures are not tailored toward maximizing revenue-related objectives such as Gross Merchandise Value (GMV), which is one of the major business metrics for e-Commerce companies. In addition, defining accurate edge relations in GNNs is non-trivial in large-scale e-Commerce systems, due to the heterogeneity nature of the item-item relationships. 
This work aims to address these issues by designing a new GNN architecture called GNN-GMVO (Graph Neural Network - Gross Merchandise Value Optimizer). This model directly optimizes GMV while considering the complex relations between items. In addition, we propose a customized edge construction method to tailor the model toward similar item recommendation task and alleviate the noisy and complex item-item relations. In our comprehensive experiments on three real-world datasets, we show higher prediction performance and expected GMV for top ranked items recommended by our model when compared with selected state-of-the-art benchmark models.
\end{abstract}

\begin{IEEEkeywords}
Recommendation Systems, Graph Neural Networks, Similar Item Recommendations, Gross Merchandise Value Optimization
\end{IEEEkeywords}

\section{Introduction}

The goal of recommender systems in e-Commerce settings is to increase the click-through rate (CTR) and revenue by recommending items that users will likely interact with and eventually purchase \cite{zheng2020price}. Similar item recommendation plays a vital role in enhancing customers’ exploration experiences on e-Commerce websites, by recommending similar substitutes for a given anchor item. This type of models enables users to be exposed to a broader set of products and allow marketing campaigns to reach potential customers effectively. Both these factors increase the chance of conversion \cite{linden2003amazon}.

Different modeling approaches are suggested in the literature to capture complex relations like similarity for items. One of the recent approaches for this task is to use graph-based models, which can identify and formulate the relation between items to make accurate recommendations \cite{wu2022graph}. GNNs generate representations of nodes that depend on the graph's structure, item pairs links (edge) features, and relations. The most common paradigm for GNNs in item recommendations is to learn node (i.e., product) representation to perform relation predictions based on the embedding vectors \cite{he2020lightgcn}.

Identifying product relations, such as similarity and complementarity, is important in the e-Commerce recommendation platform \cite{xu2020product}. Ignoring these diverse types of relations can lead to losing critical information about item relations. In addition, the ultimate optimization objective can play a critical role in graph models' performance in different settings. For instance, if the graph model's architecture optimizes relevance-related loss functions, it can lead to suboptimal recommendations from a revenue standpoint. This, in turn, can make the usage of those graph architectures prohibitive for large-scale industrial e-Commerce systems.

To address these issues, we propose a new graph-based model that considers the diverse, complex relations in item spaces of large-scale e-Commerce settings and can directly optimize on Gross Merchandise Value (GMV)-related loss function. We call this architecture Graph Neural Network - Gross Merchandise Value Optimizer (GNN-GMVO). Under this model, we propose a new multi-objective decoder function that optimizes on a combination of relevance and GMV. This makes the model capable of adjusting the loss function per the usage setting. In other words, depending on the degree of importance of relevance and revenue, the model objective can change.   

Through our extensive experiments, we design a new edge relation based on item-item data that considers relational information like co-view, view-then-bought (an item viewed, then another item is ultimately bought), and co-purchase. This new metric helps us better identify similarity relations among other types of relations between items and reduce noise (i.e., other types of item-item relations).

We perform experiments to validate the proposed architecture on a user-interaction proprietary e-Commerce dataset from \textit{Walmart.com} and two publicly available datasets from Amazon. According to the results, GNN-GMVO outperforms the currently deployed model in prediction metrics in \textit{Walmart.com}. The model also performs better than GCN \cite{kipf2016semi} and Graph Attention Networks (GAT)\cite{velivckovic2017graph} models in expected GMV without hurting the NDCG metric.    

The rest of this paper is organized as follows. In section \ref{sec:literature}, we present background and related work. In section \ref{sec:methods}, the methodology and the architecture of GNN-GMVO are elaborated. In section \ref{sec:results}, we report the experiments conducted to compare the proposed architecture with the existing benchmarks. Section \ref{sec:conclusion} states the conclusion and direction for future research.

\section{Background and related work}\label{sec:literature}

Among proposed algorithms for recommendation systems, GNNs have shown to be one of the most promising models \cite{wu2022graph}. One reason for this success might be the inherent design of the graph models that directly takes advantage of item-item, user-item, or user-sequence interactions. In the last decade, various aspects of GNN-based recommendation systems have come under attention in industry and academia \cite{darban2022ghrs, huang2002graph, tang2021dynamic}. In this section, we review some of the relevant work to our paper.

One closely related topic to similar item recommendation is social recommendation task. This task emerged with the creation of online social networks. The models in the social recommendation track assume that a given graph node's local neighbors can be used to improve node representation modeling, because a given node's neighbors should have similarity with the node itself \cite{guo2015trustsvd}. The similarities between nodes are used in two different ways for modeling purpose: (i) to improve final generated node representation modeling \cite{ma2009learning}, (ii) to explicitly use them as regularizers to limit the final node latent representations \cite{ma2008sorec, ma2011recommender, tang2013exploiting}. 

Considering diverse nature of item-item relation is one of the important aspect of our work. Because of this another stream of research pertaining to our work is the knowledge graph based recommendation. Knowledge graphs utilize a complex graph structure with several types of nodes and relation among them \cite{wang2019kgat}. Mainstream papers in this area of literature create embeddings for relations and focus on semantic relevance (see \cite{wang2019multi, zhang2018learning} for example), and the semantic information of both nodes and relations are considered.

Another research track relevant to our work is models focusing on revenue optimization. Most research in item recommendation systems' literature is based on optimizing the item recommendation relevance. However, in e-Commerce settings, the objective maybe to optimize on generated revenue \cite{yousefi2020choice}. There are non-graph machine learning models like \cite{maragheh2022prospect} and \cite{wang2018prospect} that can potentially model context created by price of recommended items on the user behavior. Some other papers take this one step further and study the so-called assortment optimization problem (see \cite{yousefi2020choice}, \cite{brovman2016optimizing}). Under assortment optimization problem, one is interested in finding a subset of items that maximize the revenues. Most of the models studied in this context, like multinomial logistic regression \cite{talluri2004revenue}, mixed multinomial logistic regression \cite{rusmevichientong2014assortment}, and nested logit model \cite{davis2014assortment} are simple listwise models from an ML standpoint of view, which makes their usage prohibitive in complex big-data settings. However, due to listwise nature of user choice behavior, the assortment optimization problem becomes combinatorially challenging (see \cite{gallego2015general} for example). For instance, \cite{brovman2016optimizing} develops a scalable model to identify similar items and maximize revenue by constructing a pointwise ranking model, see \cite{yousefi2021choice} and \cite{gallego2019revenue} for a survey of models optimizing on generated revenue. However, as mentioned, these models do not explicitly incorporate the item-item relations inherent to graph models and may lose some relational information. 

In parallel to non-graph models focusing on optimizing revenue, there are graph models for similar-item recommendation not optimizing on revenue-related metrics. However, they mainly focus on optimizing relevance in the item recommendation setting. For example, \cite{bhandari2013serendipitous} develop a method to recommend unforeseen apps to the users by constructing an app-app similarity graph and using users' interaction data with previously installed apps. \cite{qiu2019rethinking} propose an architecture with a weighted graph attention layer to provide in-session item embedding and recommend the next item in each session while optimizing CTR.

Finally, some papers attempt to present graph-based models that are at least price aware if not optimizing on generated revenue by the recommendations. In their model, the utility scores of items are function of recommended price of the whole list of recommended items. \cite{zheng2020price} propose a price-aware GNN-based recommender system that discovers user price sensitivity for each price category. To do so, they add nodes for price and category to the user-item graph and allow the item price to be propagated to the user embeddings through the item nodes. This structure allows for user price preference on unexplored categories. To the authors’ best knowledge, although revenue awareness has been a research topic for other recommender systems \cite{azaria2013movie},\cite{jannach2017price}, it has not been studied as part of GNN-based recommender systems. In this paper, we seek to fill the literature gap by proposing a graph-based model for a similar item recommendation task that explicitly models item-item relations while optimizing generated revenue by the recommendations.

\section{Method}\label{sec:methods}

This section introduces the architecture of the proposed GNN-GMVO framework, the high-level system view, and its components. In detail, we discuss GNN-GMVO model, item graph construction, and model training and inference for similar item recommendation tasks. Specifically, we focus on introducing two variants of our model built on Graph Convolutional Networks (GCNs) and Graph Attention Networks (GATs), and call them GCN-GMVO and GAT-GMVO, respectively.

\subsection{GCN-GMVO Model}\label{sec:gcn-gmvo}

Let $G=(V,E)$ denotes a graph where $V$ and $E$ represent the set of nodes and edges of the graph. Also, denote the embedding matrix of nodes by $X\in\ \mathcal{R}^{d\times|\mathcal{V}|}$, where $d$ is the dimension size of the embedding and $|\mathcal{V}|$ is the number of nodes in the graph. The graphs represent the relations defined over different sets of entities (represented by nodes). An edge $e \in E$ shows a connection between node $u$ ($u \in V$) and node $v$ ($v \in V$). 

GNNs are a class of neural network models built on graph structure and use relations defined between the nodes. GNNs aggregate information from the graph's structure to create a deep representation for each node by using a form of message passing to transfer information between nodes to update each node's representation. 

Graph Convolutional Networks (GCNs) are among the most popular GNN models. Under GCNs, message passing between nodes is done via Eq. \eqref{msg_pro}:

\begin{equation}\label{msg_pro} 
  h_u^k=\textit{ReLU}(W^k\sum_{v\in N(u) U_{\{u\}}} h_v^{k-1}/\sqrt{|N(u)||N(v)|}),
\end{equation}

\noindent where $h_u^k$ shows hidden representation of node u after $k^{\textit{th}}$ message passing step, and $W^k$ is a trainable matrix. $N(u)$ and $N(v)$ show the neighborhood nodes of $u$ and $v$, respectively. For each node, this function aggregates information from its neighborhood and combines it with the previous embedding of that node to update its representation. The input features of each node are used as initial hidden embedding $h^0$ (i.e., $h_u^0=X_u$), and after $K$ message passing steps, the final embeddings of nodes, $z_u$, are created. In other words: 

\begin{equation}\label{final_emb} 
z_u=h_{u}^K \quad \forall u\in V.
\end{equation}

GCN model operates as the encoder function by using the local graph structure around each node. The encoder maps nodes to an embedding space. A decoder function reconstructs the connections of the graph from the encoded node embeddings. In the similar item recommendation task, the decoder should perform as a predictor of the similarity between pairs of nodes in the graph. This is done by predicting whether two nodes are connected in the graph and reduces to the link prediction problem \cite{hamilton2020graph}. Under GCN, the encoder function takes the graph structure and the initial node features as the input, and generates the final embeddings of the nodes using Eq. \eqref{final_emb}. Then, a decoder function reconstructs the neighborhood structure for each node.

Under the GCN-GMVO framework for the similar item recommendation problem, we aim to optimize on item similarity relevance while inflating the weight of the edges with higher item prices. In other words, the decoder should identify similar items for a given anchor node (item) of interest, while boosting it according to the prices of respective items. To achieve this goal, we adjust the decoder function to make the final loss function more sensitive to links created among neighbors with higher prices. Because of this, the decoder function is modeled as Eq. \eqref{decoder}:

\begin{equation}\label{decoder} 
DEC\left(z_u,z_v\right)=(1+\lambda(p_u+p_v))(z_u^Tz_v),
\end{equation}
where $p_u$ is the normalized price of item $u$, and $z_u^Tz_v$ is the inner product between embeddings of nodes $u$ and $v$. Note that if $z_u^Tz_v$ is higher for a pair $(u,v)$, nodes $u$ and $v$ are more similar. Under Eq. \eqref{decoder}, the inner product of the embeddings of two nodes $u$ and $v$ are inflated by the sum of normalized prices of both nodes $u$ and $v$. $\lambda$ controls the trade-off between the importance of price and similarity. If $\lambda=0$, the decoder function only focuses on the similarity between nodes $u$ and $v$. However, by increasing the value of $\lambda$, the decoder considers more weight for the revenue generated by the corresponding nodes of the pairs.

Under GCN-GMVO, we frame the similar item recommendation problem as a link prediction problem. The link prediction is a classification problem where the positive label is assigned to a link between nodes $u$ and $v$ if an edge connects the nodes. Also, a negative link is assigned to pairs of nodes if no edge connects them in a graph structure. In order to train the model, we randomly sample a subset of positive edges and an equal number of negative edges from the training data. The sampling is done according to a uniform distribution. Then, to frame the problem as a link prediction problem, a binary cross-entropy loss (Eq. \ref{loss}), is defined on the encoder-decoder structure: 

\begin{equation}\label{loss} 
\begin{split}
\mathcal{L}= & -\frac{1}{N}\sum_{(u,v)\in E} l(DEC(z_u,z_v),A[u,v]) \\
= & -\frac{1}{N}\sum_{(u,v)\in E}{y_{(u,v)}.\log{\left(\sigma\left(DEC\left(z_u,z_v\right)\right)\right)}}+ \\
= & ({1-y_{u_n,v_n}).\log{\left(1-\sigma\left(DEC\left(z_{u_n},z_{v_n}\right)\right)\right)}}.
\end{split}
\end{equation}

In Eq. \eqref{loss}, $\sigma$ denotes the sigmoid function, which maps the decoder to a probability score, and $A$ is the binary adjacency matrix. In addition, $u_n$ and $v_n$ show the nodes for the negative samples from non-existing edges. The loss function measures the discrepancy between the decoded edge values and the true values.

\subsection{GAT-GMVO Model}\label{sec:gat-gmvo}

The decoder function defined by Eq. (\ref{decoder}) can be applied on other GNN architectures such as GAT (Graph Attention Network) \cite{velivckovic2017graph} or GraphSAGE \cite{hamilton2017inductive} to tailor these models toward optimizing revenue while decoding the edges. GAT has shown promising results by combining GNNs with attention mechanism \cite{velivckovic2017graph}. The GAT model introduces attention weight into graphs, which is used to calculate the hidden representation of the nodes by attending over neighbor’s influence during the aggregation step. Eq. (\ref{eq:attention}) shows the hidden representation of node $u$ in iteration $k$ of message passing:

\begin{equation}\label{eq:attention} 
  h_u^k=\sigma(\sum_{v\in N(u)} \alpha_{u,v} h_v^{k-1}),
\end{equation}

\noindent $\alpha_{u,v}$ shows the attention on neighbours of node $u$ while aggregating its neighbour information during the message passing. $\alpha_{u,v}$ can be calculated using Eq. ({\ref{eq:alpha}}):

\begin{equation}\label{eq:alpha} 
  \alpha_{u,v}= \frac{exp(a^{T}[Wh_u \oplus Wh_v])}{\sum_{v^\prime \in N(u)} exp(a^{T}[Wh_u \oplus Wh_{v^{\prime}}])},
\end{equation}

\noindent where $a$ shows a trainable attention vector, $W$ is a trainable matrix, and $\oplus$ denotes the concatenation operation \cite{hamilton2020graph}. Other components of GAT-GMVO framework including the decoder and loss can be modeled similar to GCN-GMVO architecture proposed in section \ref{sec:gcn-gmvo}. This change in the decoder function can make the GAT model more sensitive to highly priced neighbors. 

\subsection{Loss Function Variants}\label{sec:loss}

Cross-entropy function is used to formulate the loss function for GNN-GMVO. However, other types of ranking functions such as pairwise max margin loss can also be utilized to optimize the encoder-decoder loss for GNN-GMVO architecture. Eq. (\ref{eq:maxmargin}) shows the max margin loss function for the link prediction problem. 

\begin{equation}\label{eq:maxmargin} 
  L = \sum_{(u,v)\in E} max(0, -DEC(z_u,z_v) + DEC(z_{u_n},z_{v_n}) + \Delta),
\end{equation}

\noindent where $\Delta$ is the margin for the difference between $DEC(z_{u_n},z_{v_n})$ and $DEC(z_u,z_v)$. This loss function helps the model to optimize the weights by minimizing the scores generated by non-existing edges and maximizing the scores generated by positive edges while considering the generated revenue. Since the decoder function takes into account the revenue optimization, the loss function considers lower loss value for the edges with higher price nodes.

\subsection{Item Graph Construction, Model Training, and Model Inference}\label{sec:itemgraph}

\begin{figure*}[t!]
\centerline{\includegraphics[scale=0.45]{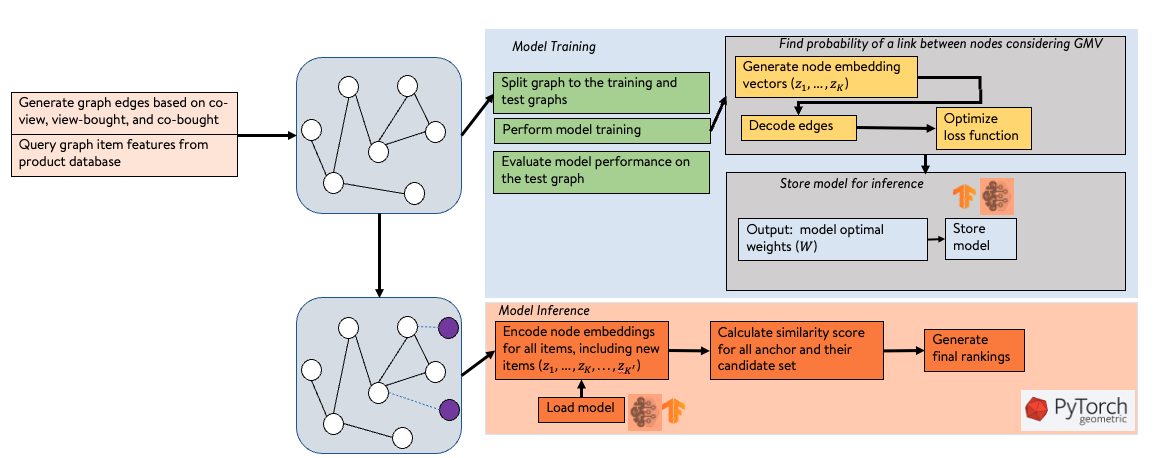}}
\caption{Proposed architecture for similar item recommendation (GNN-GMVO).}
\label{fig:arch}
\end{figure*}

Since item-item feature data in its raw format does not include edges among the items, we need to construct a graph for the training purpose to capture the similarity connections between items. To do this, we construct item features for all the nodes. This work uses a pre-trained model called Universal Sentence Encoder ($USE$) to extract initial node embeddings $X$ (see \cite{cer2018universal} and \cite{useenc}). This model creates a sentence embedding with dimensions of 512. We use the item name, item category, and other textual information of the item to generate text that is inputted to $USE$ to generate the embeddings.

In order to detect edges of the graph, we need to define a metric that represents similarity relations between items, as there could be multiple types of relations such as complementarity, substitutability, and relevance among items of a given data in e-Commerce settings. The existence of multiple types of relations can create noise in detecting similar nodes to any given item, as users' click/purchase behavior (which is used to construct the graph) can be affected by these diverse relations among the items. Hence, we use co-view, view-bought, and co-purchase data as signals to define a new and custom metric to identify similarity relations and remove noise:

\begin{equation}\label{edge} 
\begin{split}
Sc(u,v)= & |cv(u,v)|+|vb(u,v)| + \\ 
 & |vb(v,u)|-|cp(u,v)| \quad \forall u,v\in V, 
\end{split}
\end{equation}

\noindent where $\left|cv\left(u,v\right)\right|$,$\left|cp\left(u,v\right)\right| $ show the number of times items $u$ and $v$ are co-viewed and co-purchased together. $\left|vb\left(u,v\right)\right|$ shows the number of times item $u$ is viewed and then item $v$ is purchased. We subtract $\left|cp\left(u,v\right)\right|$ from co-view and view-then-bought to account for noise created by complementary items and only keep similarity relations between nodes. Note that complementary items, although bought or viewed together or after one another, are not similar. We use Eq. \eqref{threshold_edge} to detect strong similarity relations while removing noise created by complementary items:

\begin{equation}\label{threshold_edge}
edge(u,v)=
    \begin{cases}
        1 & \text{if } Sc(u,v)>\theta\\
        0 & \text{ } otherwise.
    \end{cases}
\end{equation}
Eq. \eqref{threshold_edge} assumes an edge between $u$ and $v$ if $Sc(u,v)$ is larger than a threshold $\theta$.

We randomly split the graph into training and test sets to train the model and evaluate its performance. The edges are partitioned between training and test sets. We use a two-hop message passing GCN and GAT models with an output embedding of 256 dimensions. In other words, Eq. (\ref{msg_pro}) and  Eq. (\ref{eq:attention}) run for two iterations to encode the nodes into final node embeddings $z_u$, $\forall u \in \mathcal{V}$.
Then given sampled negative and positive edges, the loss function \eqref{loss} is obtained. After completion of training, the obtained model is used for inferencing and ranking.

To perform the model inference on a set of items, we consider all those items as nodes of the graph.
We assume an edge between the nodes of a given pair if Eq. \eqref{edge} holds for that pair. The inference graph is encoded as mentioned in subsections \ref{sec:gcn-gmvo} and \ref{sec:gat-gmvo} using the trained model. See subsection \ref{coldstart} for the proposed approach of constructing the edges for cold or new items with zero or small number of views and transactions.

\subsection{Ranking Task for Recommendation}\label{sec:ranking}

In this paper, we only focus on optimizing revenue and relevance via using the graph models. Therefore, in order to retrieve similar items, one can use a similarity measure between two items' final graph embeddings. After model inference and encoding nodes to embedding space, we apply a weighted similarity function to find the similarity score between nodes $u$ and $v$ using Eq. \eqref{cosine}:
\begin{equation}\label{cosine} 
score\left(u,v\right)=\ (1+\lambda(p_u+p_v))\times(\frac{z_u^Tz_v}{|z_u||z_v|}).
\end{equation}

\noindent Eq. \eqref{cosine} represents the weighted cosine similarity score between item $u$ and item $v$. When ranking recall set for an anchor item $u$, the higher the price of item $v$ ($v \in$ \textit{Candidate set for u}), the higher it will be ranked.

\subsection{Edge Construction for Cold Items}\label{coldstart}

In a large-scale e-Commerce implementation of any recommendation module, new items with no user-interaction history are always added. In addition, some items may have low user traffic, and they may not be connected to any other graph nodes using a threshold-base rule like Eq. \eqref{edge}. The new and low-traffic (cold) items become isolated in the graph in these cases. Therefore, the node embeddings generated by the GNN-GMVO model will only use the weights of that node, and no message will be passed from the other nodes of the graph:

\begin{equation}
  h_u^k=\textit{ReLU}(W^k h_u^{k-1}).
\end{equation}

In these cases, one may ignore a threshold-based rule like Eq. \eqref{edge} and add a connection from the node with the most similar initial embedding, $h_0$, to the new/cold item. This case may alleviate the isolation problem of the new/cold items, connect them to the rest of the graph, and obtain better final embeddings. The items with the highest probability are more likely to be similar to each other: 

\begin{equation}\label{cold_Start} 
{S}_i=\textit{argmax}(\frac{ e^{X_i X_j}} {\sum_{j^\prime \in V} e^{X_i X_j^\prime}}) \quad \forall j^\prime\in V. 
\end{equation}
$S_i$ shows the most similar item to cold item $i$. For graphs with millions of nodes, approximate nearest neighbour search algorithms such as \cite{johnson2019billion} and \cite{malkov2018efficient} could be utilized to retrieve the most similar nodes. Figure \ref{fig:arch} summarizes the architecture of the GNN-GMVO model and the steps explained in this section.

\section{Results}\label{sec:results}
\subsection{Results on Walmart dataset}

\begin{figure*}[t]
\centerline{\includegraphics[scale=0.30]{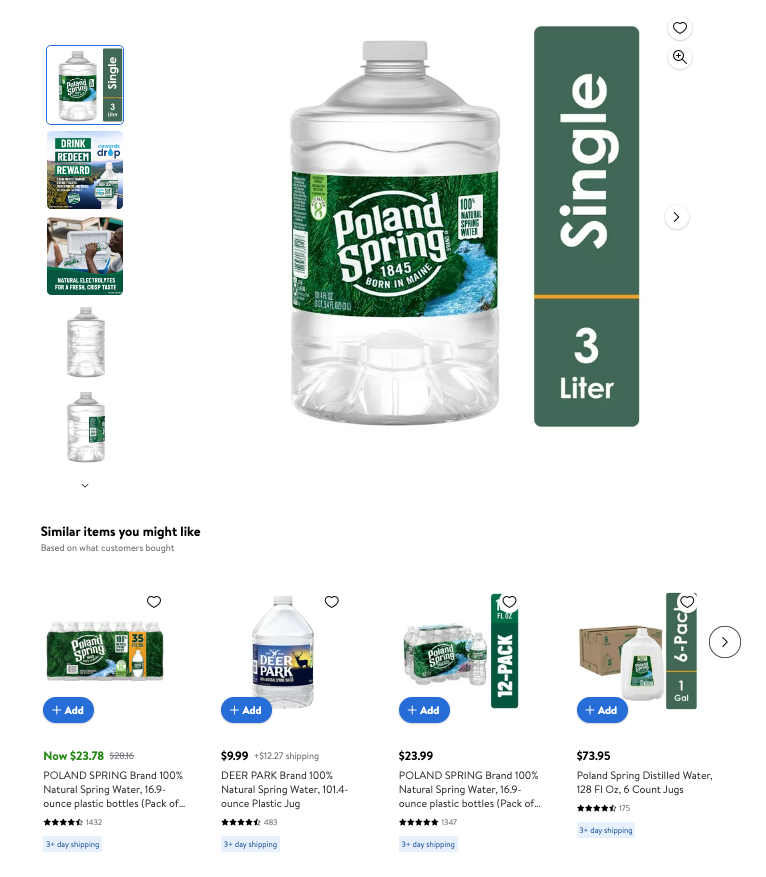}}
\caption{Similar item recommendation module on \textit{Walmart.com} item pages.}
\label{fig:si}
\end{figure*}

This section summarizes the conducted empirical experiments to validate the proposed modeling approach on real data instances. We compare the model with the currently deployed recommendation architecture in the similar item recommendation module of \textit{Walmart.com}. We call this  benchmark model which is a sophisticated Deep Learning model $SIRB$ (Similar Item Recommendation Benchmark) in the rest of this paper. The similar item recommendation module of \textit{Walmart.com} is in the item pages of the platform and seeks to recommend alternative similar items to the main item of the page (See Figure \ref{fig:si}).

We use Normalized Distributed Cumulative Gain (NDCG) to measure the ranking relevance. We also define a metric to measure the expected generated GMV by the models' suggested rankings ($EGMV@K$). To calculate $EGMV@K$, after the model ranking, the first $K$ items are determined for each anchor item. Then, we use the transaction data obtained for a month after model inferencing to approximate the ground truth. We calculate the portion of times each candidate item is being transacted and consider this as the probability of the purchase for each item. More specifically, $EGMV@K$ is calculated as follows:

\begin{equation}
GMV@K=\sum_{k=1}^{K}\sum_{t=1}^{T}{\frac{|Tr_{t,k}|}{\sum_{j\in c_a}{|Tr_{t,j}|}}P_k}, 
\end{equation}

\noindent where $|Tr_{t,k}|$ is the number of times candidate item $k$ is purchased at time $t$, $\sum_{j\in c_a}{|Tr_{t,j}|}$ shows the total number of transactions for all candidate items under an anchor item $a$, and $P_k$ shows price for candidate item $k$.

The data is proprietary and is 
selected from the Online Grocery category of items and includes about 100,000 items and 4 million links (as the edges of the graph) among those items. The graph edges are constructed using Eq. \eqref{threshold_edge}.

We conducted parameter tuning by training the model with different learning rates (0.001, 0.01, 0.1) and epoch sizes (10, 20, 50, and 100). We also tested the model performance using one, two, and three-hop message passing parameter. The optimal inference results are achieved using learning rate of 0.1 for the ADAM optimizer, the epoch size of 20 with two-hop message passing. PyTorch Geometric 2.2.0 is used for training and inference. Each epoch takes approximately 30 seconds to finish using a machine with 32GB memory and 10 CPU cores. We train our proposed model with different values of $\lambda$ in Eq. \eqref{decoder} to find $\lambda$ that maximizes the objective function. When inferencing, node embeddings are generated by loading the trained model and passing the inference graph as a dataset to encode nodes and generate graph embeddings. The inference dataset includes about 50,000 nodes. The candidate set for each anchor item is ranked based on the weighted cosine similarity of the recommended items and the given anchor item (i.e., Eq.\eqref{cosine}). The results are presented in Table 1. Most of the users at item pages of \textit{Walmart.com} view at most two sets of four items when checking similar item recommendations. This is why we report @8 metrics. 

We conducted ablation study to show the importance of the proposed decoder function in optimizing the objective function. When $\lambda=0$, the model is equivalent to a traditional GCN model without a GMV optimizer. As can be seen in Table \ref{table:walmart_result}, using GCN model with custom edges defined in Eq. \eqref{edge} increases NDCG metric by $4.2\%$ w.r.t. the benchmark SIRB (when $\lambda=0$). By increasing $\lambda$ in GCN-GMVO, the $NDCG$ metric decreases for $\lambda\geq1$. This is expected since for larger values of $\lambda$ the loss becomes more focused on optimizing the revenue. In addition, from Table \ref{table:walmart_result}, we see that increasing $\lambda$ up to 0.5 increases the $EGMV@8$ metric, but larger $\lambda$ has an inverse effect as it biases the loss too much so that items relevance (and ultimately $EGMV@8$) becomes small. Setting $\lambda=0.1$ achieves both high $NDCG$ and $EGMV$ scores, which translates to increasing both relevance and expected GMV in our test sets. This proves the efficiency of our modeling framework in optimizing both recommendation relevance and revenue in large-scale e-Commerce settings.

\begin{figure*}[t!]
     \centering
     \begin{subfigure}[b]{0.45\linewidth}
         \centering
         \includegraphics[width=\linewidth]{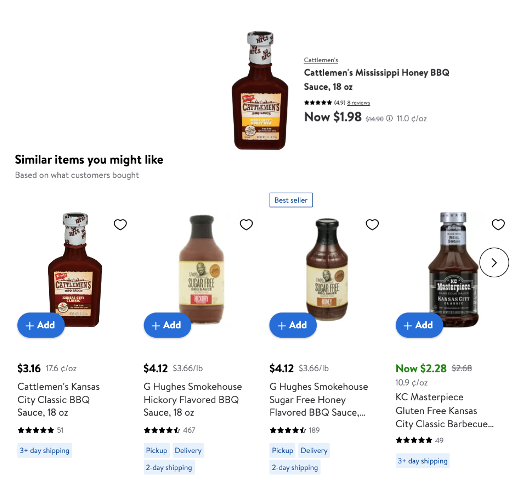}
         \caption{SIRB top-4 recommendations for an anchor item}
         \label{fig:sirb_reco}
     \end{subfigure}
     \begin{subfigure}[b]{0.45\textwidth}
         \centering
         \includegraphics[width=\linewidth]{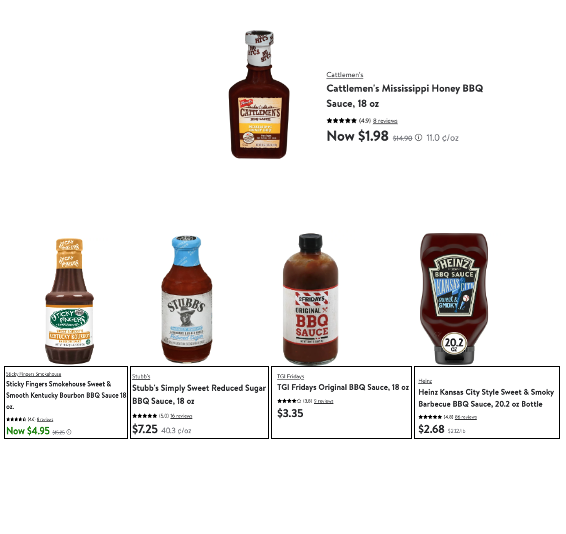}
         \caption{GCN-GMVO top-4 recommendations for an anchor item}
         \label{fig:gcn_reco}
     \end{subfigure}
    \caption{Comparison between items recommended by SIRB and GCN-GMVO}
    \label{fig:comparison}
\end{figure*}

Figure \ref{fig:comparison} shows the set of recommended items for a grocery anchor item by SIRB and GCN-GMVO. The results show that GCN-GMVO model recommends different set of items compared to the SIRB algorithm. The recommended items could potentially decrease the total number of sales if they are too expensive. However, since GNN-GMVO algorithm controls the trade-off between relevance and revenue, by optimizing the loss function we can improve $EGMV$ generated by the recommendation set. As can be seen from Figure \ref{fig:gcn_reco}, the price of the recommendations for the anchor item are slightly more than the recommendations of SIRB (Figure \ref{fig:sirb_reco}). However, the model shows some degree of price elasticity which results in $EGMV$ boost without hurting $NDCG$ metric. This is an example of how the proposed algorithm could positively impact the item recommendation task for large-Scale e-Commerce platforms.

\begin{table*}[t!]
\centering
\caption{NDCG  and  EGMV  scores  for  GCN-GMVO and SIRB recommendation algorithms for \textit{Walmart.com} dataset.}
\begin{tabular}{||l l l||} 
 \hline
 $\lambda$ & GCN-GMVO ($NDCG@8$, $EGMV@8$)  & SIRB ($NDCG@8$, $EGMV@8$) \\ [1ex] 
 \hline\hline
 0 & (0.3041 , 3.0253) & (0.2918 , 2.8848)  \\ 
 0.05 & (0.3050 , 3.0332) & -  \\
 0.1 & \textbf{(0.3060 , 3.0372)} & - \\
 0.5 & (0.3054 , 3.0194) & -\\
 1 & 0.2927 , 2.8664) & - \\
 2 & (0.2995 , 2.9905)  & -\\ [1ex] 
 \hline
\end{tabular}
\label{table:walmart_result}
\end{table*}

\begin{table*}[tb]
\centering
\caption{NDCG  and  EGMV for publicly available datasets.}
\begin{tabular}{||l l l||} 
\hline
\multicolumn{3}{c}{Panel A: NDCG  and  EGMV  scores  for  GCN and GAT algorithms for All-Beauty Category.}\\ \hline 
$\lambda$ & GCN-GMVO ($NDCG@4$, $EGMV@4$)  & GAT-GMVO ($NDCG@4$, $EGMV@4$) \\ [1ex] 
 \hline\hline
 0 & (0.641 , 1.95) & (0.644 , 1.94)  \\ 
 0.05 & (0.641 , 1.96) & (0.644 , 1.95)  \\
 0.1 & (0.641 , 1.97) & (0.644 , 1.96) \\
 0.5 & (0.641 , 1.98) & (0.644 , 1.96)\\
 \textbf{0.8} & \textbf{(0.641 , 2.02)} & \textbf{(0.644 , 1.98)} \\
 1 & (0.641 , 1.97) & (0.644 , 1.95)\\
 2 & (0.641 , 1.96) & (0.644 , 1.96)\\ [1ex] 
 \hline
 \hline
 \multicolumn{3}{c}{Panel B: NDCG  and  EGMV  scores  for  GCN and GAT algorithms for Video Game Category.}\\ \hline 
 $\lambda$ & GCN-GMVO ($NDCG@4$, $EGMV@4$)  & GAT-GMVO ($NDCG@4$, $EGMV@4$) \\ [1ex] 
 \hline\hline
 0 & (0.312 , 6.37) & (0.323 , 6.67)  \\ 
 0.05 & (0.309 , 6.29) & \textbf{(0.327 , 6.76)}  \\
 0.1 & (0.314 , 6.43) & (0.327 , 6.74) \\
 0.5 & (0.305 ,  6.22) & (0.321 , 6.63)\\
 0.8 & (0.313 , 6.48) & (0.318 , 6.55) \\
 1 & (0.313 , 6.47) & (0.323 , 6.69)\\
 2 & \textbf{(0.314 , 6.49)} & (0.323 , 6.71)\\
 3 & (0.309 , 6.31) & (0.319 ,  6.57)\\[1ex] 
 \hline
\end{tabular}
\label{table:amazon_result}
\end{table*}

\subsection{Results on Publicly Available Dataset}

We also evaluate the performance of our architecture on two separate categories (All Beauty and Video Games) of Amazon datasets  \cite{ni2019justifying, data}.

\subsubsection{All Beauty Category}

The dataset contains metadata from Amazon for 32,992 products in the "All Beauty" Category. Product title, description, price, also\_bought (list of items bought after viewing the product), also\_viewed (list of items viewed after viewing the product), brand, and category features are used in this experiment. 
Some parts of the data set are missing. For instance, 65\% of the products do not have price data. Because of this, we compute the price average and standard deviation to fill in missing values by sampling from the price distribution. The initial node representations are computed by inputting the item's textual information (including category, description, title, and brand) to the USE encoder \cite{cer2018universal, useenc}. Since the frequency for the items co-view, view-bought, and co-purchase are missing in the dataset, we connect the product pairs in the graph if they are co-viewed even once. 
In other words, in order to build the graph, $\theta$ in Eq. \eqref{threshold_edge} is considered to be $0$ for this experiment.

We train our proposed architecture with different values of $\lambda$ in Eq. (\ref{decoder}). The candidate set for each anchor item is based on the items that are viewed after the anchor item. We rank recall set based on the weighted cosine similarity of the recommended items and the given anchor items using Eq. (\ref{cosine}). In order to evaluate the models, we measure and compare their performance on top-4 recommended items based on the total number of transactions for each item. 

Traditional GCN and Graph Attention Networks (GAT) are considered as the benchmark models for this experiment. The cross-entropy loss function converges after around 20 epochs when training these models. The total run time with a 32GB memory is approximately 5 minutes. We used the same training configuration of the \textit{Walmart.com} dataset in this experiment. The results show improvement in GMV without impacting the NDCG scores of the recommendations for some cases of positive $\lambda$s. One of the goals of the proposed architecture is to improve revenue, consequently GMV, in the large-scale e-Commerce systems without recommending irrelevant items compared to the anchor item. Both of these goals are measured by comparing $NDCG$ and $EGMV$ of the recommendation sets.   

As can be seen from Table \ref{table:amazon_result}-Panel A, when $\lambda=0$ 
$EGMV@4$ and $NDCG@4$ for GCN are 1.95 and 0.641, respectively. They are 1.94 and 0.644 for GAT model. Setting with $\lambda=0.8$ generates $EGMV@4=2.02$ for GCN-GMVO and $EGMV@4=1.98$ for GAT-GMVO models which yields 3.6\% and 1.6\% improvements in $EGMV@4$ w.r.t. GCN and GAT benchmark models. Using GCN and GAT architectures along with GMV optimizer as the decoder function can improve GMV metric without hurting the relevance of the recommendations. However, the results show that the GCN is performing slightly better than GAT model in optimizing revenue in this experiment.

\subsubsection{Video Games Category}

This dataset contains 84,893 items from video game category. We used the same model configuration and same set of features as the previous experiment to evaluate the model performance on a bigger item graph. We used the same pre-trained model (i.e., $USE$) to generate item text embedding using item categories, description, title, and brand features. Similar to the previous experiment, the results show that our architecture can improve $EGMV$ generated by the recommendation system. We observe that the GAT model with GMVO component outperforms GCN model in optimizing the revenue. 

The results (Table \ref{table:amazon_result}-Panel B) show that the GAT-GMVO model has the highest $EGMV$ when $\lambda=0.05$. This shows that the optimal trade-off between relevance and revenue happen when we consider lower weights for the nodes’ price. However, we see optimal $EGMV$ for the GCN-GMVO model when $\lambda=2$. As can be seen from the table, $EGMV$ is improved by 1.3\% for the GAT-GMVO.  This improvement in the $EGMV$ has no negative impact on the NDCG of the recommendations. We also see that $EGMV$ is improved by 1.8\% under GAT-GMVO model.

\section{Conclusion}\label{sec:conclusion}

In this paper, we propose Graph Neural Network-Gross Merchandise Value Optimizer (GNN-GMVO) architecture to optimize GMV while considering complex item-item relations for similar item recommendation task. We develop a new decoder to adjust the generated loss function to become more sensitive to the price of the recommended items. 
We define a new edge construction framework in the item graph to identify similarity relations between items and remove noise caused by other types of relations in the item space. We propose a step-by step framework to (i) query input feature data, (ii) construct the graph, (iii) train the model, and (iv) use it for inference and ranking tasks. We conduct extensive experiments to train the model and find the optimal weights to trade off GMV and relevance. 
The results show that the proposed model improves the expected GMV on three real datasets without hurting the NDCG scores of the recommendations. This may prove the usefulness of the proposed approach in further optimizing the GMV in some industrial recommendation settings.
In future work, one may add more features such as items’ features (price, product type, product category, etc.) and image embedding data to the textual embedding data to enrich the initial embedding inputs of the model and test if this makes the modeling framework more robust.
Finally, feeding generated node embeddings by GNN-GMVO architecture to other expert models like wide \& deep(W\&D) \cite{cheng2016wide}, prospect-net \cite{maragheh2022prospect}, DeepFM \cite{guo2017deepfm}, and xDeepFM \cite{lian2018xdeepfm} can be tested and examine if this further optimizes the model.

\bibliography{main}
\bibliographystyle{IEEEtran}

\end{document}